\documentclass[emulateapj]{article}
\usepackage{emulateapj}

\begin{document}

\title{The Lack of Very Ultraluminous X-ray Sources in Early-type Galaxies}

\author{Jimmy A. Irwin\altaffilmark{1}, Joel N. Bregman\altaffilmark{1},
and Alex E. Athey\altaffilmark{2}}

\altaffiltext{1}{Department of Astronomy, University of Michigan,
Dennison Building, Ann Arbor, MI 48109; jairwin@umich.edu, jbregman@umich.edu}

\altaffiltext{2}{Observatories of the Carnegie Institute of Washington,
813 Santa Barbara St, Pasadena, CA 91101; alex@ociw.edu}

\begin{abstract}
We have searched for ultraluminous X-ray sources (ULXs) in a sample of 28
elliptical and S0 galaxies observed with {\it Chandra}. We find that the number
of X-ray sources detected at a flux level that would correspond to a 0.3--10 keV
X-ray luminosity of $\sim2 \times 10^{39}$ ergs s$^{-1}$ or greater (for which
we have used the designation very ultraluminous X-ray sources; VULXs) at the
distance of each galaxy is equal to the number of expected foreground/background
objects. In addition, the VULXs are uniformly distributed over the
{\it Chandra} field of view rather than distributed like the optical light of
the galaxies, strengthening the argument that the high flux sources are
unassociated with the galaxies. We have also taken the VULX candidate list
of Colbert and Ptak and determined the spatial distribution of VULXs in
early-type galaxies and late-type galaxies separately. While the spiral
galaxy VULXs are clearly concentrated toward the centers of the galaxies,
the early-type galaxy VULXs are distributed randomly over the
{\it ROSAT} HRI field of view, again indicating that they are not associated
with the galaxies themselves. We conclude that with the exception of
two rare high luminosity objects within globular clusters of the
elliptical galaxy NGC~1399, VULXs are generally not found within old stellar
systems. However, we do find a significant population of sources with
luminosities of $1-2 \times 10^{39}$ ergs s$^{-1}$ that reside within
the sample galaxies that can be explained by accretion onto 10--20 M$_\odot$
black holes. Given our results, we propose that ULXs be defined as X-ray
sources with $L_X$(0.3--10 keV) $> 2 \times 10^{39}$ ergs s$^{-1}$.

\end{abstract}


\section{Introduction}
The existence of very luminous X-ray point sources located within galaxies
but not coincident with the galaxy's nucleus has spawned a flurry of
investigations into the nature of these objects in recent years. Usually defined
as having an X-ray luminosity exceeding $10^{39}$ ergs s$^{-1}$,
these ultraluminous X-ray objects (ULXs) were originally discovered in the
{\it Einstein} era (Fabbiano 1989), and studied further with {\it ROSAT}
and {\it ASCA} (e.g., Colbert \& Ptak 2002; Makishima et al.\ 2000;
Roberts \& Warwick 2000). The excellent spatial resolution of {\it Chandra}
has led to the discovery of many more ULXs densely packed within
star-forming regions of interacting galaxies such as the Antennae
(Zezas \& Fabbiano 2002).

While a few ULXs have been identified with supernovae, the majority of ULXs
are thought to be accreting objects owing to the fact that their X-ray flux
can vary significantly over timescales of months and years.
The X-ray luminosities of ULXs can exceed the Eddington luminosity of a
1.4 M$_{\odot}$ neutron star by up to two orders of magnitudes, calling into
question the physical nature of these sources. One possible
explanation set forth to explain the origin of ULXs are intermediate
mass (50--1000 M$_{\odot}$) black holes accreting at or near their
Eddington limit (Colbert \& Mushotzky 1999). Such a population of intermediate
mass black holes would represent a missing link between stellar-mass black
holes like those found in the Milky Way and supermassive black holes at
the centers of galaxies. However, it is uncertain how a black hole in this
mass range can form, and arguments both for (Miller \& Coleman 2002;
Miller et al.\ 2003) and against (King et al.\ 2001; Roberts et al.\ 2002;
Begelman 2002)
this scenario have been presented previously. Another possible explanation is
that ULXs are normal accreting low-mass black holes for which the X-ray emission
is beamed toward us (King et al.\ 2001). In such a model, King  et al.\ (2001)
proposed that the most likely candidate in a beaming scenario is a
phase of thermal time scale mass transfer in binaries with intermediate or
high mass donor stars. The need for an intermediate or high mass donor
star naturally explains the tendency for ULXs to be found within galaxies
with recent strong star-formation.

\begin{table*}[t]
\footnotesize
\caption{Sample of Galaxies}
\label{tab:sample}
\begin{center}
\begin{tabular}{lccrlccrlcc}
\multicolumn{11}{c}{} \cr
\tableline \tableline
Galaxy & Obs. & Distance &\vline& Galaxy & Obs. & Distance &\vline& Galaxy
& Obs. & Distance\cr
&ID&(Mpc)&\vline&&ID&(Mpc) & \vline&&ID&(Mpc) \cr
\tableline

NGC~1316& 2022& 21.5& \vline &NGC~3923 & 1563& 22.9 &\vline &NGC~4494 & 2079& 17.1\\
NGC~1332& 4372& 22.9& \vline &NGC~4125 & 2071& 23.9 &\vline &NGC~4552 & 2072& 15.4\\
NGC~1399 & 319 & 20.0& \vline &NGC~4261 &  834& 31.6 &\vline &NGC~4621 & 2068& 18.3\\
NGC~1404 & 2942& 21.0& \vline &NGC~4365 & 2015& 20.4 &\vline &NGC~4636 & 323 & 14.7\\
NGC~1549 & 2077& 19.7& \vline &NGC~4374 & 803 & 18.4 &\vline &NGC~4649 & 785 & 16.8\\
NGC~1553 & 783 & 18.5& \vline &NGC~4382 & 2016& 18.5 &\vline &NGC~4697 & 784 & 11.8\\
NGC~3115 & 2040&  9.7& \vline &NGC~4406 &  318& 17.1 &\vline &NGC~5846 & 788 & 24.9\\
NGC~3377 & 2934& 11.2& \vline &NGC~4459 & 2927& 16.1 &\vline &IC~1459  & 2196& 29.2\\
NGC~3379 & 1587& 10.6& \vline &NGC~4472 & 321 & 16.3 &\vline &&&\\
NGC~3585 & 2078& 20.0& \vline &NGC~4486 & 2707& 16.1 &\vline &&&\\

\tableline
\end{tabular}
\end{center}
\end{table*}

Such a beaming scenario would argue that ULXs should not be found within old
stellar populations such as S0 and elliptical galaxies. However, a recent
compilation of ULX candidates in nearby galaxies by Colbert \& Ptak (2002)
found 34 ULX candidates within a sample of 15 elliptical and S0 galaxies.
While Colbert \& Ptak (2002) caution that some of the ULX candidates might be
unrelated foreground/background objects, the extent to which contamination
from unrelated sources might affect the ULX rate within early-type galaxies
was not determined. Indeed, Irwin, Athey, \& Bregman (2003) found that the
number of very high X-ray flux sources within a sample of 13 elliptical and S0
galaxies observed with {\it Chandra} was exactly that expected from
contaminating foreground/background sources. While there was an overabundance
of sources with X-ray luminosities in the $1-2 \times 10^{39}$
ergs s$^{-1}$ luminosity range, the number of sources with apparent
luminosities exceeding $2 \times 10^{39}$ ergs s$^{-1}$ in the 13 galaxies
was exactly what was expected from foreground/background objects.

Still, only ten very high X-ray flux sources in the sample of
13 galaxies limited the strength of the claim that early-type systems
do not generally harbor very luminous ULXs.
In this {\it Letter} we present data from a
larger sample of galaxies observed with {\it Chandra} to verify that very
luminous ULXs (VULXs),
defined as sources with 0.3--10 keV luminosities exceeding
$2 \times 10^{39}$ ergs s$^{-1}$, are absent from elliptical and S0 galaxies.

\section{Observations and Data Reduction} \label{sec:observations}

For our sample we began with all elliptical and S0 galaxies that were
available in the {\it Chandra} archive as of October 2003 for which
the galaxy was observed with the S3 chip of the ACIS detector. We have
only included galaxies such that a $10^{39}$ ergs s$^{-1}$
source would contain at least 36 source counts (or four times the standard
detection limit) so that incompleteness would not be an issue. This limited
us to galaxies within $\sim35$ Mpc for a typical {\it Chandra} observation.
We excluded the very nearby early-type galaxies NGC~5102 and NGC~5128 for
which a visual inspection of the images revealed no high flux sources,
and for which the expected number of foreground/background high flux sources
was too small to make their addition to the sample worthwhile. These
criteria led to a sample of 30 galaxies.

Determining an accurate distance to each galaxy is crucial if we hope to
ascertain whether sources more luminous that a certain limit are present in
the sample or not. We have chosen to use the $I$-band surface brightness
fluctuation distances ($I$-SBF) of Tonry et al.\ (2001),
since this study provides the
largest homogeneously-analyzed sample of all the distance estimator methods,
and includes all 30 galaxies in our sample. For each galaxy, we searched
for an independent estimate of the distance, typically from the globular cluster
luminosity function method or the $K$-band surface brightness fluctuation
method, and in general found good agreement. Two exceptions were NGC~1407
and NGC~720, for which the $I$-SBF distances were 60\% and 30\% larger
than other distance estimators, respectively (Perrett et al.\ 1997;
Mei et al.\ 2001). Given the large uncertainties to these galaxies,
we have excluded them for our sample. For the other galaxies we have assumed
the $I$-SBF distances, which are consistent with distances derived from
other methods at the 15\% level. The 28 galaxies are listed in
Table~\ref{tab:sample}.

The galaxies were processed in a uniform manner following the
{\it Chandra} data reduction threads.
The data were calibrated with the most recent gain maps
at the time of reduction.
Pile-up was not an issue even for the brightest sources and no
correction has been applied.
Sources were detected using the ``Mexican-Hat" wavelet detection routine
{\sc wavdetect} in CIAO in an 0.3--6.0 keV band image.

Count rates were converted into fluxes using a $\Gamma=2$ power
law model, a value typical for high flux sources in early-type galaxies
(Irwin et al.\ 2003). We also assumed Galactic hydrogen column densities
from Dickey \& Lockman (1990). We considered fitting the spectrum of each
individual ULX in order to derive a flux for each source, but our main
goal was to compare the number of detected high flux sources to the number
of sources predicted from deep field studies. Since previous deep field studies
have assumed a single power law model to convert counts to flux, we have
done the same. All X-ray luminosities are quoted in the 0.3--10 keV range
unless noted otherwise.

Depending on the angular size of the galaxy, from five to 16 effective
(half-light)
radii would fit at least partially on the ACIS-S chip. For each ULX candidate,
its projected position within the galaxy as a function of effective radii was
noted (e.g., between 0--1 effective radii, between 1--2 effective radii,
etc.), where the effective radius for each galaxy has been collected from
the literature. This was done in order to create a combined
radial profile for the ULX candidates to determine their spatial distribution.
In a future paper, we will catalog the positions,
count rates, spectral parameters, and other relevant information of the
ULX candidates presented here.

\section{Results} \label{sec:results}

\subsection{The Number of Expected Foreground/Background Sources}
\label{subsec:expected}

To determine the number of unrelated foreground/background X-ray sources
expected in the field of view of each observation, the
log $N$ vs.\ log $S_X$ relation derived by Hasinger et al.\ (1998) was employed.
We considered using one of the {\it Chandra} deep field studies (i.e.,
Giacconi et al.\ 2001), but the number of very high
flux sources found within a typical {\it Chandra} field of view is so small
that the slope of the log $N$ vs.\ log $S_X$ relation at the high flux end
is poorly constrained, and would lead to poorly constrained
values for the expected number of high flux sources. Conversely, the
{\it ROSAT} HRI-derived log $N$ vs.\ log $S_X$ relation of
Hasinger et al.\ (1998) was determined from both a deep observation of the
Lockman Hole and many shorter pointed observations of isolated, non-extended
X-ray sources in order to sufficiently populate the high flux end of the
log $N$ vs.\ log $S_X$ relation.

The log $N$ vs.\ log $S_X$ relation given in Hasinger et al.\ (1998) is
(upon integrating the differential form):

\begin{equation} \label{eq:xlum}
N(>S_X) = \left\{
\begin{array}{l}
117.0~S_X^{-0.94} - 20.1 \qquad S_X < 2.66 \\
\\
138.4~S_X^{-1.72} + 0.87 \qquad S_X > 2.66 \\
\end{array}
\right. \, ,
\end{equation}
where $N(>S_X$) is the number of sources with fluxes exceeding $S_X$
per square degree, and 
$S_X$ is the 0.5--2.0 keV flux of the source in units of $10^{-14}$
ergs s$^{-1}$ cm$^{-2}$. For each of the 28 {\it Chandra} fields, we calculated
the number of expected background/foreground sources with 0.5--2.0 keV fluxes
corresponding to 0.3--10 keV luminosities of $1-2\times 10^{39}$ ergs s$^{-1}$
and $>2\times 10^{39}$ ergs s$^{-1}$ (if the sources are assumed to be at the
distance of the target galaxy). In total, 33.1 and 33.5
sources were expected in the 28 {\it Chandra} fields that would be mistaken
as $1-2\times 10^{39}$ ergs s$^{-1}$ and $>2\times 10^{39}$ ergs s$^{-1}$
sources, respectively. The number of such sources actually detected was
$75^{+9.7}_{-8.6}$ and $31^{+6.6}_{-5.5}$, respectively, where the errors
given are $1\sigma$ Poisson uncertainties.
So while there is a significant excess in the
number of $1-2\times 10^{39}$ ergs s$^{-1}$ sources over what was expected,
the number of $>2\times 10^{39}$ ergs s$^{-1}$ sources detected is precisely
what is predicted from deep field studies.

\subsection{The Spatial Distribution of the High Flux Sources} 
\label{subsec:spatial}

In addition to comparing the number of foreground/background
sources expected to the number of ULX candidates in the galaxies, the
distribution of the sources can provide an independent estimate of the
amount of contamination from foreground/background objects. We created
one combined radial distribution profile of all the sources in the sample,
as discussed above. In order to improve the statistics, the 16 original radial
bins were grouped into five larger bins: 0--1, 1--2, 2--3, 3--7, and 7--16
effective radii. The number of ULX candidates in each bin was divided by
the area of the bin in units of the effective area. This
was done so that a random distribution of sources would yield a flat radial
profile, while a distribution that followed the optical light of the
galaxy would be significantly peaked toward the center.
The combined radial profile is shown in Figure~\ref{fig:radial}.
We have plotted the
distributions for the $1-2\times 10^{39}$ ergs s$^{-1}$
and $>2\times 10^{39}$ ergs s$^{-1}$ sources separately.
At large radii, the profiles of both luminosity
classes are flat, indicative of a distribution that is randomly distributed.
Within two effective radii, there is a clear peak in the profile of
the $1-2\times 10^{39}$ ergs s$^{-1}$ sources. Of the 19 sources found within
one effective radius, only 1.6 are expected to be foreground/background
contaminants, indicating that most of the sources belong to the galaxies
in the sample. Such a strong peak is not seen in the radial profile of the
$>2\times 10^{39}$ ergs s$^{-1}$ sources. While there is a slight
overabundance of $>2\times 10^{39}$ ergs s$^{-1}$ sources within one
effective radius (four were detected while 1.7 were expected), this
overabundance is only significant at the 90\% confidence level. Furthermore,
of the four $>2\times 10^{39}$ ergs s$^{-1}$ sources within one effective
radius, three have luminosities below $2.4\times 10^{39}$ ergs s$^{-1}$.
Thus, even if these three sources do belong to the galaxies they are only
minimally above the $2\times 10^{39}$ ergs s$^{-1}$ threshold, and would
fall below this threshold if their host galaxies were only 8\% closer
than was assumed. Finally, for two of the three sources, fitting their
spectra (rather than assuming a $\Gamma=2$ spectral model) yielded luminosities
less than $2.2\times 10^{39}$ ergs s$^{-1}$ for these two sources.
\centerline{\null}
\vskip3.10truein
\includegraphics{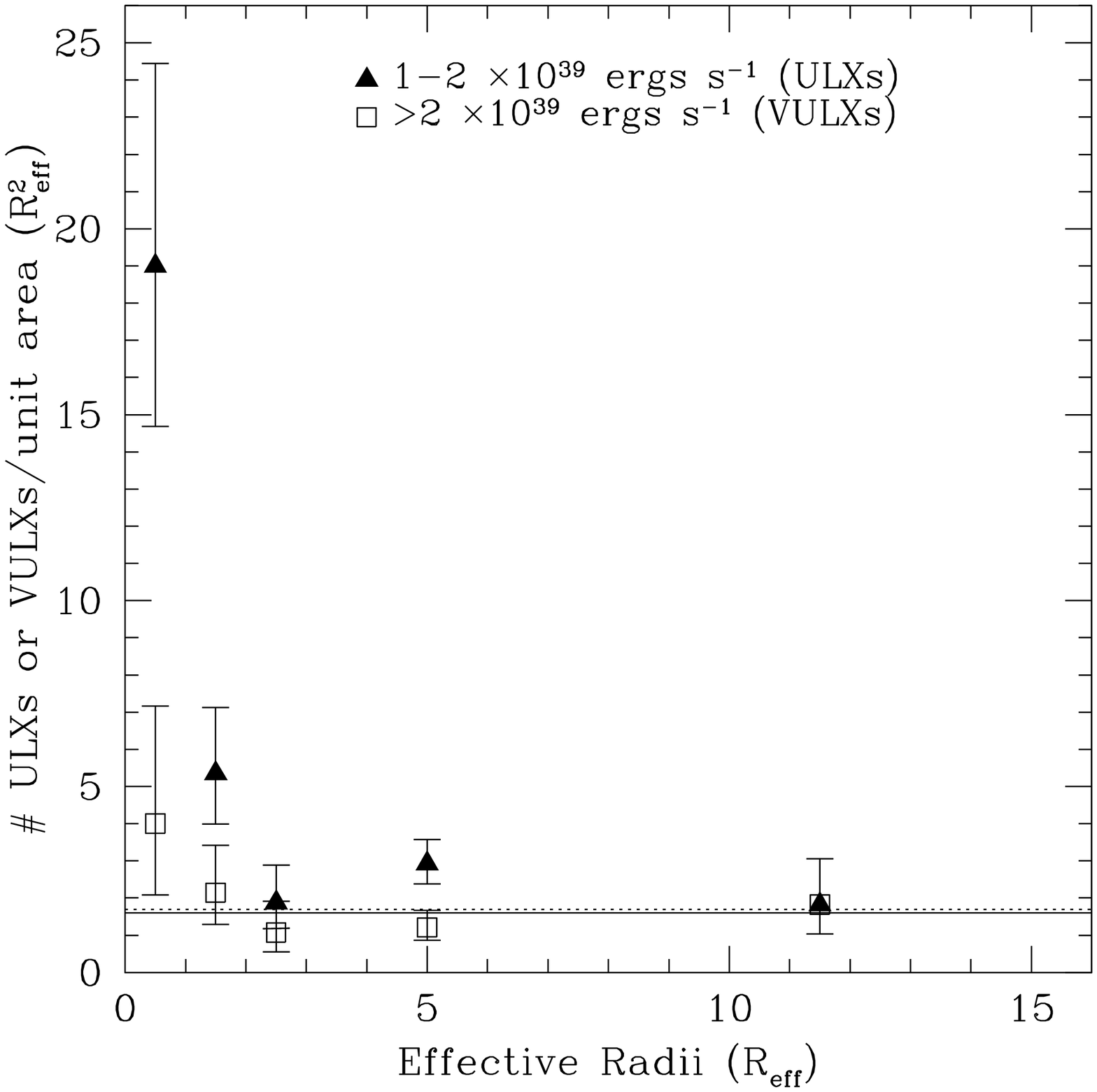}
\figcaption{ \small
Composite radial profiles of sources with apparent luminosities of
$1-2\times 10^{39}$ ergs s$^{-1}$ ({\it filled triangles}) and
$>2\times 10^{39}$ ergs s$^{-1}$ sources ({\it open squares}) normalized by
the area of each spatial bin for all sources within the 28 galaxies of our
sample. Errors are $1\sigma$. The solid and dotted lines represent the expected
radial distribution of unrelated background/foreground sources for the two
luminosity classes.
\label{fig:radial}}
\normalsize

\subsection{Comparison to the Colbert \& Ptak (2002) VULX Catalog}
\label{subsec:colbert}

Colbert \& Ptak (2002) analyzed 15 early-type and 39 late-type galaxies
observed with the {\it ROSAT} HRI and found a total of 87 ULX candidates.
There is some overlap between the early-type galaxies in their sample
and ours, and in general they surveyed out to approximately the same
radial distance as we have. Colbert \& Ptak (2002) scaled
the galaxies by $R_{25}$ rather than the effective radius ($R_{25}$ is
typically 3--5 times larger than the effective radius), and included in
their catalog all sources with 2--10 keV luminosities greater than
$10^{39}$ ergs s$^{-1}$ within twice the $R_{25}$ contour (or approximately
6--10 effective radii).
Since not all of the observations were complete down to 
$10^{39}$ ergs s$^{-1}$, it is not
possible to compare the expected number of foreground/background sources
to the number of ULX candidates found, but it is possible to look at
the spatial distribution of the ULX candidates.
Colbert \& Ptak (2002) assumed a power law of $\Gamma=1.7$ rather
than the $\Gamma=2.0$ model used in our work. After accounting for the
difference in spectral model choice and energy band, it was found that 
the lower luminosity limit of their catalog was $L_X$ (0.3--10 keV) $\sim 
2 \times 10^{39}$ ergs s$^{-1}$ (they are all what we would call VULXs).
Based on the results of
\S~\ref{subsec:expected} and \S~\ref{subsec:spatial}, we would therefore
expect the distribution of their sources to be consistent with a random
distribution.

From Table~1 of Colbert \& Ptak (2002), we determined how many VULX
candidates were within radial bins of 0--0.5 $R_{25}$, 0.5--1.0 $R_{25}$,
1.0--1.5 $R_{25}$, and 1.5--2.0 $R_{25}$. We included any high X-ray flux
QSOs that were excluded by Colbert \& Ptak (2002), and excluded a VULX
candidate in NGC~4374 (IXO 50) for which {\it Chandra} images showed it
to be a clump of hot gas rather than a point source. We also excluded from the
sample any galaxy for which twice $R_{25}$ exceeded $17^{\prime}$ since
the outer radial bins would not fit within the {\it ROSAT} HRI field of view.

\centerline{\null}
\vskip3.10truein
\includegraphics{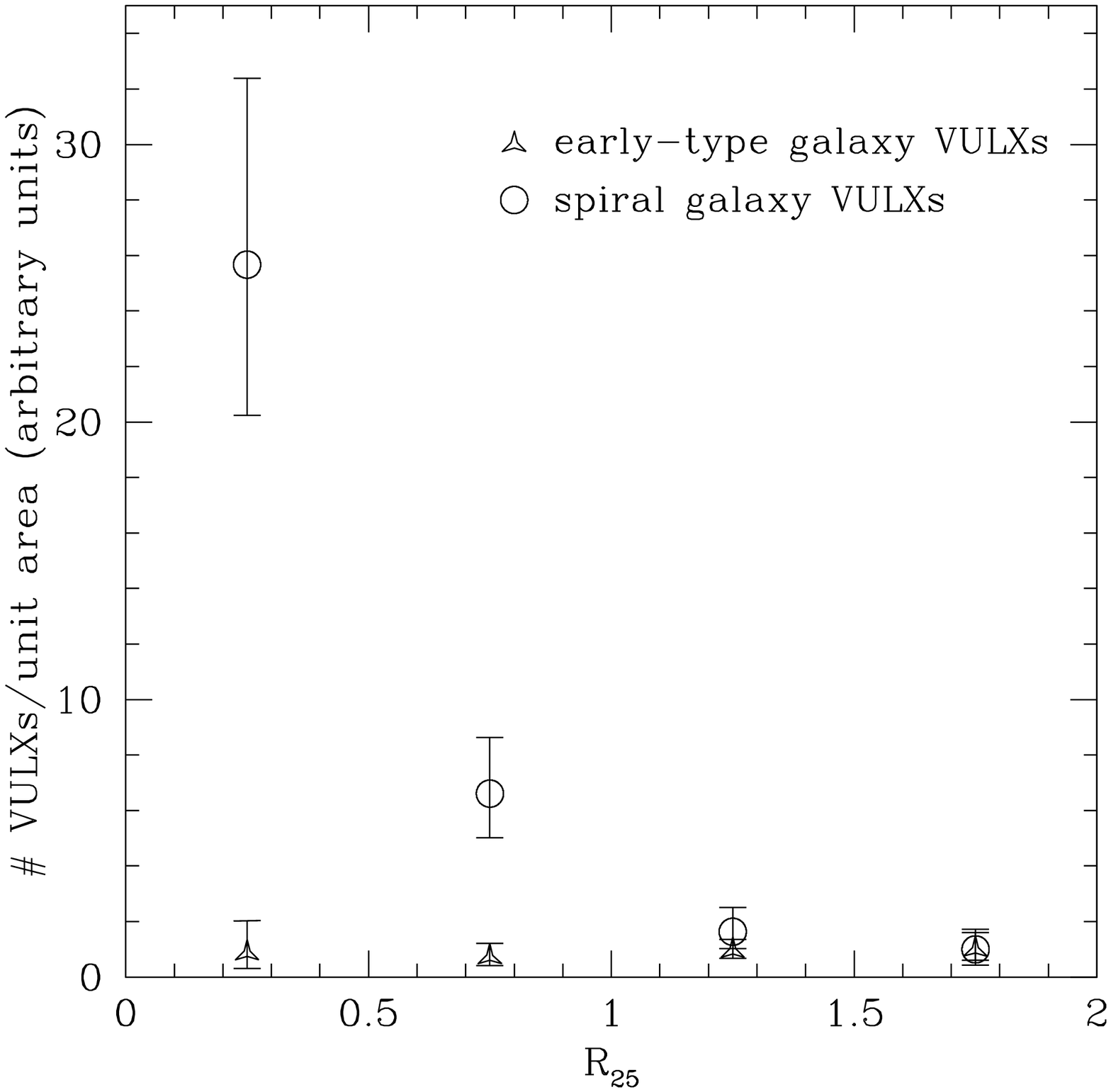}
\figcaption{ \small
Composite radial profiles of VULXs compiled by Colbert \& Ptak (2002)
for early-type galaxies and spiral galaxies
normalized by the area of each spatial bin. The bins
have been renormalized such that the fourth spatial bin of each profile has
a value of 1.0.
Errors are $1\sigma$.
\label{fig:colbert}}
\normalsize
\vspace{0.1truein}
For a random distribution, we would expect the number of sources detected
within each radial bin to scale simply with the ratio of the areas of the
four bins. For 0--0.5 $R_{25}$, 0.5--1.0 $R_{25}$, 1.0--1.5 $R_{25}$,
and 1.5--2.0 $R_{25}$ bins, this would be a 1:3:5:7 ratio. The number of
VULX candidates found in these four bins were 2, 5, 11, and 16, respectively.
Within the Poisson uncertainties, the distribution of the VULX
candidates are completely consistent with being random (and therefore
unrelated to the galaxies). Interestingly, a similar exercise with
late-type galaxies in the Colbert \& Ptak (2002) catalog yields a completely
different result; within the four radial bins there were 22, 17, 7, and
6 VULX candidates, respectively. The sources are heavily concentrated
toward the centers of the galaxies, indicating that nearly all of the
sources within $R_{25}$ are bona fide VULXs. This is not surprising given
the well-known connection between recent star formation and the presence
of VULXs. The results for both the early-type and spiral galaxy samples
are illustrated in Figure~\ref{fig:colbert}, where the VULX distributions
for the two samples have both been normalized to 1.0 in the fourth spatial
bin for easier comparison.

\subsection{Two VULXs Within Globular Clusters of NGC~1399}
\label{subsec:globulars}

Angelini, Loewenstein, \& Mushotzky (2001) found that two sources with
$L_X > 2 \times 10^{39}$ ergs s$^{-1}$ were coincident with globular
cluster candidates, which would conflict with the idea that sources
this luminous avoid early-type systems. We located these two sources
in the {\it Chandra} data of NGC~1399. The spectrum of one of the sources
was very soft ($\Gamma=2.5$), and with this spectral model we determined
its X-ray luminosity to be $2.3~(d/20~{\rm Mpc})^2 \times 10^{39}$ 
ergs s$^{-1}$. Thus, if NGC~1399 is only 8\% closer (within the
uncertainties of the Tonry et al.\ 2001 distance estimate), the luminosity of
this source would drop below $2 \times 10^{39}$ ergs s$^{-1}$. The other
source has a considerably higher luminosity
($L_X= 4.7 \times 10^{39}$ ergs s$^{-1}$) and cannot be explained
by assuming a slightly smaller distance. It is possible that the optical
counterpart is really a background AGN that has been miss-identified as
a globular cluster. We have analyzed the same {\it HST} data that
Angelini et al.\ (2001) used to obtain a globular cluster list for this galaxy,
and have verified that the object has a globular cluster-like $B-I$ color
and apparent magnitude. However, at the distance of NGC~1399, a globular cluster
would be at best marginally-resolved by {\it HST}, so only a high resolution
spectrum of the optical counterpart will unambiguously determine if it is
a globular cluster or an AGN. At any rate, given the large number of
globular clusters that NGC~1399 contains, as well as the large number of
globular clusters in our sample as a whole, we can at least conclude that
$> 2 \times 10^{39}$ ergs s$^{-1}$ sources are extremely rare in globular
clusters, and in early-type systems in general.

\section{A Standard Definition for an Ultraluminous X-ray Source}
\label{sec:definition}

There does not appear to be a standard definition for a ULX in the literature.
Various studies have defined ULXs differently, using different luminosity
thresholds and different energy bands. Here, we propose that a ULX
be defined as having a 0.3--10 keV luminosity that exceeds
$2 \times 10^{39}$ ergs s$^{-1}$ (what we have been calling VULXs in this
paper). Using $2 \times 10^{39}$ ergs s$^{-1}$
as a break between ULXs and normal X-ray binaries has two advantages.
First, we note that X-ray sources that have luminosities of
$1-2 \times 10^{39}$ ergs s$^{-1}$ can be adequately explained by
accretion onto a 10--20 M$_{\odot}$ black hole (for which observational
evidence already exists for blacks holes of this mass in our own Galaxy,
e.g, Cygnus X-1), thus eliminating the need for a more
exotic explanation such as beaming or the existence of intermediate mass
black holes.
Second, $>2 \times 10^{39}$ ergs s$^{-1}$
sources are apparently lacking (or at least very rare) in old stellar
populations, suggesting that an additional class of object is present in
late-type galaxies that is absent in early-type galaxies.

\acknowledgements
J. A. I. was supported by NASA grant G02-3110X, and J. N. B. acknowledges
support from NASA grant NAG5-10765. We thank the referee, Ed Colbert, for
many useful suggestions that improved the manuscript.

\end{document}